\documentclass{ws-procs975x65}

\newcommand{\qa}{Q_{\! A}}
\newcommand{\qm}{Q_{\! M}}

\begin{document}

\title{\uppercase{Quintessence with Kaluza-Klein type couplings to matter and an isotropy-violating vector field}\footnote{Based on work done in collaboration with Sigbj\o rn Hervik and David F. Mota.\cite{thorsrud12}}}
\author{\uppercase{Mikjel Thorsrud}}
\address{Institute of Theoretical Astrophysics, University of Oslo, N-0315 Oslo, Norway}




\begin{abstract}
We study the dynamics of a scalar field with Kaluza-Klein type couplings to cold dark matter and an isotropy-violating vector field. The vector coupling, $f^2(\phi)F^2$, has been studied thoroughly in the context of inflation recently. We generalize the model to a dark energy context and study the cosmological consequences. We find a rich set of exact anisotropic power-law solutions and identify a strong vector coupling regime where the anisotropy is controllable and all solutions are close to the $\Lambda$CDM limit.   
\end{abstract}
\bodymatter

\section*{Introduction}
Scalar fields provide a simple framework for 
exploring 
the current 
acceleration of 
the universe in terms of a dynamical  dark energy component.  If not forbidden by an unknown symmetry, scalar fields are expected to couple to other matter fields \cite{carroll98}.  The coupling to cold dark matter, via a conformal rescaling of the matter sector metric, has been studied extensively in the literature; we shall refer to the scenario proposed in Ref. \refcite{amendola99} as Standard Coupled Quintessence (SCQ). Another natural possibility, which has been studied thoroughly in the context of inflation recently\cite{soda12},  is the coupling to a vector field through the $f^2(\phi)F^2$ mechanism. This coupling preserves the gauge invariance, is ghost free and leads to a stable, but small anisotropy in the expansion of the universe.  Higher dimensional theories, such as string theories, however, quite typically predict both the matter and vector coupling in the effective action \cite{gasperini01}; the original Kaluza-Klein theory is historically the first example.  Although both couplings are common in fundamental theories, the vector coupling has mainly been ignored in the context of late time cosmology. 
With a particular emphasize on the vector field's back-reaction on geometry, we summarize the cosmological consequences of a scalar field simultaneously coupled to cold dark matter and an isotropy-violating vector field. We name the model Doubly Coupled Quintessence (DCQ) and refer to our recent paper for details.\cite{thorsrud12} 
\section*{Dynamical system}
In the Einstein frame the model is summarized in the following Lagrangian density: 
\begin{equation}
	\mathcal{L} = \frac{1}{2}R - \frac{1}{2}g^{\mu\nu} \nabla_\mu \phi \nabla_\nu \phi - V_0 e^{\lambda \phi} - \frac{1}{4} e^{2\qa\phi}F_{\mu\nu}F^{\mu\nu} + \mathcal L_M(\tilde g_{\mu\nu}, \Psi_i),
\label{lagrangian}
\end{equation}
where $R$ is the Ricci scalar of the metric $g_{\mu\nu}$, $\phi$ is the canonical scalar field responsible for the acceleration of the universe, $F_{\mu\nu} = \nabla_\mu A_\nu - \nabla_\nu A_\mu$ is the field strength of the vector $A^\mu$ and $\Psi_i$ represents other matter fields conformally coupled to $\phi$ via $\tilde g_{\mu\nu}=e^{2 \qm\phi} g_{\mu\nu}$.  The exponential type functions, which parametrize the scalar field's potential and couplings, are motivated by dimensional reduction. $\lambda$, $\qa$ and $\qm$ are constant parameters controlling the shape of the potential and the strength of the couplings to $A^{\mu}$ and $\Psi_i$, respectively.  We leave these parameters free from the start and later identify the cosmologically interesting region of parameter space. We assume a homogenous vector pointing in the x direction so that $\mathbf{A}\equiv A_\mu dx^\mu = A(t)dx$ in the gauge $A_0=0$. To model a realistic cosmology, we let $\Psi_i$ account for both cold dark matter and the cosmic microwave background; both represented as perfect fluids. 
The simplest spacetime consistent with the symmetries of the matter sector is the locally rotationally symmetric Bianchi type I metric:
\begin{equation}
ds^2 = -dt^2 + e^{2\alpha(t)}\left[  e^{-4\beta(t)}dx^2 + e^{2\beta(t)} (dy^2 + dz^2)   \right].
\label{metric}
\end{equation}
We define a dimensionless shear variable $\Sigma\equiv (d\beta/dt)/(d\alpha/dt)$.
With these assumptions the model reduces to the scenario studied for inflation in the special case $\Psi_i \rightarrow 0$ \cite{kanno10,dimopoulos10,hervik11}. In the case of a vanishing vector ($A^{\mu} \rightarrow 0$) and FLRW metric ($\Sigma\rightarrow0$) it reduces to SCQ.\cite{amendola99} We shall see that the coexistence of all fields leads to genuinely new behavior absent in both of the aforementioned special cases.  The field equations can be written as an autonomous system of dimensionless variables obeying
\begin{equation}
1=\Sigma^2 +  \Omega_\text{kin}+  \Omega_V + \Omega_A + \Omega_m + \Omega_r,
\end{equation}
where $\Omega_i$ is the Hubble normalized energy densities associated with the scalar field's kinetic term ($\Omega_\text{kin}$), scalar field's potential ($\Omega_{V}$), vector ($\Omega_A$), dark matter ($\Omega_m$) and radiation  ($\Omega_r$).  Using standard methods we identify a rich set of exact power-law solutions which are expressed in terms of the free parameters ($\lambda$,$\qa$,$\qm$). A subset of these are the well known isotropic solutions of SCQ, while others are genuinely new anisotropic scaling solutions, see Ref. \refcite{thorsrud12} for complete specifications. 
\section*{Cosmology}
Our main effort was to identify a region of parameter space ($\lambda$,$\qa$,$\qm$) connecting the exact solutions into viable cosmological trajectories.  We identified a strong vector coupling regime 
\begin{equation}
|\qa|\gg 1, \quad |\qa|\gg |\qm|, \quad |\qa| \gg \lambda, 
\label{strongvector}
\end{equation}
where only a subset of the solutions are cosmologically relevant. The standard solution $\Omega_r=1$ is the only way to realize a cosmologically viable radiation dominated epoch.  The matter and dark energy dominated epochs, however, can be realized by both isotropic or anisotropic scaling solutions.  In the strong vector coupling regime there are essentially four qualitatively different ways to realize the sequence $(\text{radiation domination})\rightarrow(\text{matter domination})\rightarrow (\text{dark energy domination})$:
\begin{equation} 
 \Big( \text{isotropic} \Big) \quad \rightarrow \quad 
\left(
\begin{array}{ccc} \text{anisotropic}  \\  \text{or} \\ \text{isotropic} 
\end{array} 
\right) 
\quad \rightarrow \quad
\left(
\begin{array}{ccc} \text{anisotropic}  \\  \text{or} \\ \text{isotropic} 
\end{array} 
\right). 
\end{equation}
Which sequence that will be realized is understood from the stability and existence conditions for the exact solutions, which depend on ($\lambda$,$\qa$,$\qm$). Each of the sequences are equally probable in the sense that they occupy roughly equally large regions of parameter space. The richness of it's dynamical solutions clearly distinguish DCQ from other anisotropic universe models in the literature.  For instance, the sequence involving anisotropization during the transition from radiation to matter domination, followed by isotropization during the transition to dark energy domination, represents genuinely new behavior, see \fref{fig1}(a). Unlike other models with a simple stationary matter coupling, it is also possible to realize a matter dominated epoch followed by a late time attractor mimicking the present universe with $30\%$ matter and $70\%$ dark energy, see \fref{fig1}(b). There is a well-defined $\Lambda$CDM limit $\qm/\qa\rightarrow 0$ and $\lambda/\qa\rightarrow 0$ in which the anisotropic scaling solutions go to the standard $\Lambda$CDM solutions. The expansion anisotropy, which is known to be small observationally, is therefore controllable. We show that the modified temperature pattern of the cosmic microwave background is within observational constraints when $|\qm/\qa|\lesssim 10^{-5}$ and $|\lambda/\qa|\lesssim 10^{-4}$.  The implications of our model for inhomogeneous perturbations remain to be studied and is expected to constrain the viable region of ($\lambda$,$\qa$,$\qm$) further.
\def\figsubcap#1{\par\noindent\centering\footnotesize(#1)}
\begin{figure}[b]%
\begin{center}
  \parbox{2.3in}{\epsfig{figure=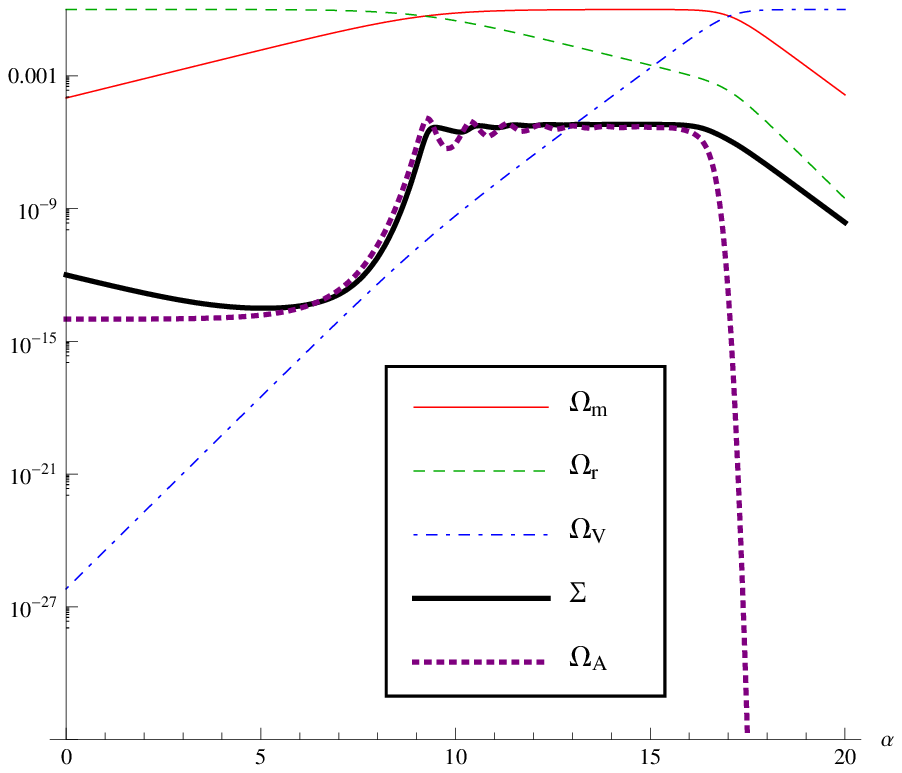,width=2.1in}\figsubcap{a}}
  \hspace*{4pt}
  \parbox{2.3in}{\epsfig{figure=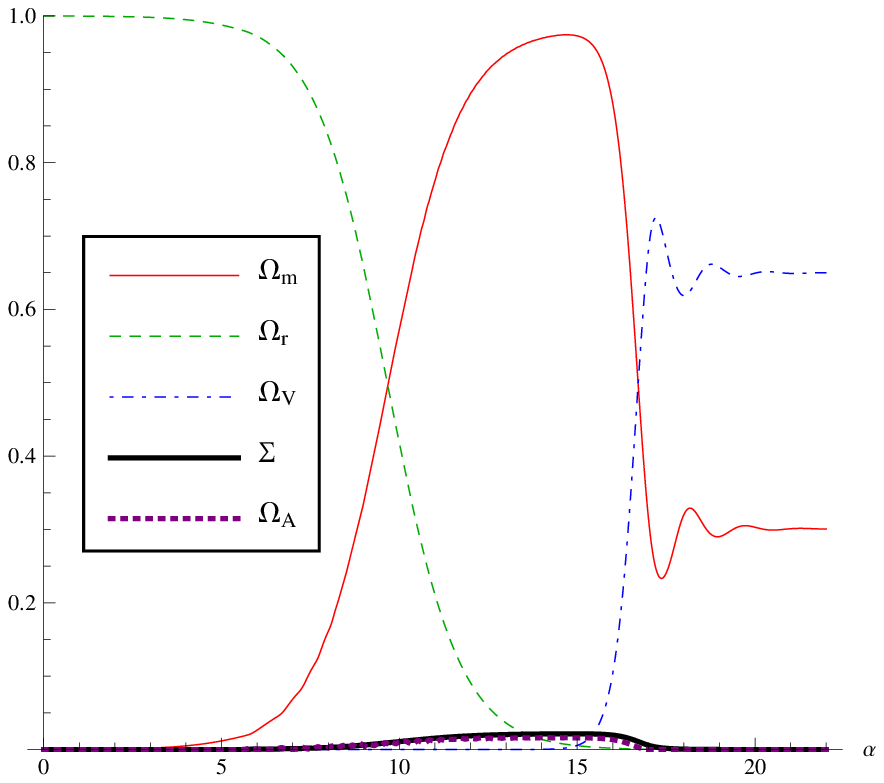,width=2.1in}\figsubcap{b}}
   \caption{Cosmological evolution for (a) ($\lambda$,$\qa$,$\qm$)$=(0.1,-10^3,-10^{-2})$ and  (b) $(2.2,-10^2,-3.3)$.}
  \label{fig1}
\end{center}
\end{figure}



\bibliographystyle{ws-procs975x65}
\bibliography{ws-pro-sample}

\end{document}